\documentclass[12pt]{article}
\usepackage{epsf, cite, amsmath, amssymb}
\usepackage{epsfig}

\makeatletter
\renewcommand\section{\@startsection {section}{1}{\z@}%
                                   {-3.5ex \@plus -1ex \@minus -.2ex}%
                                   {2.3ex \@plus.2ex}%
                                   {\normalfont\large\bfseries}}
\renewcommand\subsection{\@startsection{subsection}{2}{\z@}%
                                     {-3.25ex\@plus -1ex \@minus -.2ex}%
                                     {1.5ex \@plus .2ex}%
                                     {\normalfont\bfseries}}
\makeatother

\newcommand{\bea}{\begin{eqnarray}}
\newcommand{\eea}{\end{eqnarray}}
\newcommand{\be}{\begin{equation}}
\newcommand{\ee}{\end{equation}}

\newcommand{\Z}{{\mathbb Z}}
\newcommand{\R}{{\mathbb R}}
\newcommand{\M}{{\cal M}}

\newcommand{\Spin}{\operatorname{Spin}}
\newcommand{\SO}{\operatorname{SO}}
\renewcommand{\O}{\operatorname{O}}

\begin{document}
\begin{titlepage}

\begin{center}
September 25, 2001
%\today
\hfill                  hep-th/0109197

\hfill DUKE-CGTP-01-12, EFI-01-36

\vskip 2 cm
{\Large \bf Novel Type I Compactifications}\\
\vskip 1.25 cm 
{
David R. Morrison$^1$ and Savdeep Sethi$^2$}\\
\vskip 0.5cm
{\sl 
{}$^1$ Department of Mathematics, Box 90320, Duke University, Durham, NC
27708, USA \\ 
\vskip 0.2cm
{}$^2$ Enrico Fermi Institute, University of Chicago, Chicago, IL
60637, USA\\}

\end{center}

\vskip 2 cm

\begin{abstract}

\baselineskip=18pt

We argue that there are two distinct classes of type I
compactification to four dimensions on any space. These two classes
are distinguished in a mysterious way by the presence (or absence) of
a discrete 6-form potential. In simple examples, duality suggests that the new
class of compactifications have reduced numbers of moduli. 
We also point out analogous discrete
choices in M, F and type II compactifications, including some with
$G_2$ holonomy. These choices often result in spaces
with frozen singularities.

\end{abstract}

\end{titlepage}

\pagestyle{plain}
\baselineskip=18pt

\section{Introduction}

The moduli space of string compactifications is an intricate and complex object. Even 
compactifications with sixteen supersymmetries contain surprising
amounts of new 
physics~\cite{triples}. 
The goal of this letter is to describe new classes of type I
compactifications to four dimensions, or lower. These
compactifications are quite mysterious because they involve discrete
choices that are very hard to measure. This makes it difficult to
analyze the low-energy physics, and leaves much to be understood. 
However, we shall see that combinations of
dualities do suggest that these new compactifications exist, and that
they generically have reduced numbers of moduli.

There are also discrete choices in type II, M and F theory
compactifications to four dimensions or lower, which we also decribe. 
The geometries involved in these compactifications typically contain 
``frozen singularities'' generalizing
the kind encountered in \cite{Schwarz:1995bj, Landsteiner:1998ei, Witten:1998bs, triples}. 
Our study of discrete choices is by no
means complete. Rather, the particular discrete choices that we
describe have a natural origin in orientifold physics.

The basic mechanism that we want to consider is most easily described
in compactifications to three dimensions. These compactifications
typically have a number of abelian gauge fields. Let us take one such
gauge field, $A$. In three
dimensions, we can dualize this gauge field to obtain
a scalar field $\phi$,
\begin{equation}
d \phi = \ast F,
\end{equation}
where $F=dA$ is the field strength. However, this scalar field
actually has a compact moduli space; it takes values in $U(1)$. To
see this, we can partially rewrite the action in term of $\phi$:
\begin{eqnarray}
\frac{1}{g^2} \int F \wedge \ast F & = & \frac{1}{g^2} \int F \wedge d\phi
= \frac{1}{g^2} \int d\left( F \, \phi \right) \\
& = & \frac{1}{g^2} \int_{S^2} F \langle \phi\rangle .
\end{eqnarray} 
We denote the zero-mode expectation value of $\phi$ by $ \langle
\phi\rangle $. The 
integral of $F$ over $S^2$ measures the first Chern class of the
bundle and so is integer. Shifting $\phi$ by $2\pi g^2$ therefore has no effect
on the path-integral, and leaves the physical theory unchanged. 

What
will  concern us is
how the physics depends on the modulus $ \langle \phi\rangle $. In
perturbation 
theory, nothing depends on $\langle \phi\rangle $ because the shift symmetry
requires that only derivatives of $\phi$ appear in any low-energy
effective action. 
However, non-perturbative instanton corrections do depend on $\langle
\phi\rangle $ since 
the $n$-instanton action is weighted by a factor, 
\begin{equation}
\label{weight}
 e^{i n \langle \phi\rangle / g^2}.
\end{equation}   
Although the abelian theory is free and so trivial, we could also
imagine starting with a non-abelian theory, for example,
N=8 $SU(2)$ Yang-Mills. On the Coulomb branch of this theory, we are again
left with a single light photon. However, the moduli space is now $
(\R^7 \times S^1 )/ \Z_2$ because there are 7 scalars transforming in
the adjoint, and we have to mod out by the Weyl group. What is key for
us is that there are two fixed points on $S^1$ at $0$ and
$\pi$. Indeed, this very system appears in the study of $D2$-branes
and $O2$-planes, where the
physics at the two fixed points can be very different~\cite{sethiO2,
seiberg16, Sethi:1997sw, Banks:1997my};
see also~\cite{Berkooz:1998sn, kolhan}. 
This is a germane example in field theory of the kind of phenomena that we shall
encounter in string theory. Namely, a discrete modulus which is
undetectable in perturbation theory does, nevertheless, strongly alter
the low-energy physics. 

In a similar way, we can consider a string moving in four
dimensions. The string couples to a $2$-form gauge-field $B$ with
field strength $H=dB$. We can dualize $H$ to get a scalar, 
$$ d\phi = \ast H. $$
That this scalar is also $U(1)$-valued follows 
by essentially the same argument given above after we note that
$$ \int_{S^3} H \in \Z.$$
This compact scalar is always present in any compactification of closed
string theory from ten dimensions to four, with constant dilaton. We can also understand the
appearance of this scalar from the ten-dimensional perspective. If we
dualize $H$ in ten dimensions, we obtain a $6$-form gauge-field,
$A_6$, 
coupling to the NS $5$-brane,
$$ dA_6 = \ast H. $$
{}From this perspective, the
expectation value $\langle \phi\rangle $ corresponds to 
$$ \int_{\M_6} A_6, $$
where $\M_6$ is the compactification space.

As an example, consider type IIB
string theory compactified on $T^6$. There are $70$ scalars parametrizing the
moduli space $E_{7(7)}/SU(8)$. At least two of these scalars -- those
coming from dualizing the NS-NS and RR $2$-forms -- arise in
the manner we have described. String perturbation theory is
insensitive to the values of these compact moduli. Wrapped Euclidean $5$-branes,
however, will contribute to amplitudes in the four-dimensional
effective action. Each of these BPS instanton contributions is weighted by
a factor of the form appearing in equation (\ref{weight}). For a discussion of
such effects, see~\cite{Obers:1999um, Obers:2000ta}.
We can also consider higher dimensional theories. For example, the $3$-form
potential $C_3$ coupling to a membrane moving in five dimensions is
dual to a compact
scalar. This modulus is always present in compactifications of M
theory to five dimensions. 

Typically, in cases of low or no
supersymmetry, potentials will be generated for these moduli. For most
of our discussion, this should not play a role. 

\section{Novel Orientifolds}

The first case of interest to us is that of string theory compactified
to four dimensions on a space $\M_6$ with volume $V_6$. Since we will
be dualizing various interactions, we measure this
volume with respect to the Einstein frame metric. Recall that type I
string theory can be viewed as an orientifold of type IIB
by world-sheet parity, $\Omega$. Under the action of $\Omega$, the NS-NS
B-field, RR 0-form $C_0$, and 4-form $C_4$ are inverted 
\begin{equation}
B_2 \rightarrow - B_2,  \quad C_0 \rightarrow - C_0, \quad C_4 \rightarrow - C_4.
\end{equation}
However, this leaves a discrete choice of turning on a half-unit of
$B_2$ through a $2$-cycle and a half-unit of $C_4$ through a
$4$-cycle. The $B_2$ flux results in a type I compactification without
vector structure~\cite{sensethi, Bianchi:1992eu}. The case of $C_4$ flux
for a $T^4$ compactification has been recently studied in~\cite{Keurentjes:2001cp}.     

Since $B_2$ is inverted, its dual potential $A_6$ is also inverted by $\Omega$. We
therefore have the freedom of turning on a half-unit of $A_6$ flux through
$\M_6$.  Therefore, there are $2$ classes of type I compactification to
four dimensions. From a purely four-dimensional perspective, we have a
compact scalar $\phi$ dual to $B_2$. Normalizing $\phi$ to be dimension
one with conventional kinetic terms, we see that\footnote{Modulo factors of
$2$ and $\pi$.}
$$ \phi \sim \phi + \frac{\alpha'}{\sqrt{V_6}}. $$
Under the $\Z_2$ projection, there are two invariant values: $0$ and $\frac{\alpha'}{2\sqrt{V_6}}$. 
The first is a conventional type I compactification; the second is a
mysterious new choice. 

We might worry that we generate an anomaly for this second
choice. However, string perturbation theory is completely insensitive
to this discrete modulus so any such anomaly would have to be very
subtle. We will see in the case of $T^6$ that an anomaly is unlikely. 

\subsection{Toroidal compactification}

Is there really any difference in the low-energy physics? This is a
natural second concern. Let us turn to the case of compactification on
$T^6$. We will use a combination of dualities to convert this
compactification into something less mysterious. Our starting point is
IIB on $T^6/\Omega$ with $A_6$ flux through the torus. If we T-dualize
along $T^6$, the $A_6$ potential remains invariant. 

This might seem strange even for T-duality on one circle which gives type IIA on
$$\left( T^5 \times S^1/\Z_2 \right)/\Omega,$$
or type I'.  At first, it appears that we have converted 
a $\Z_2$-valued field into a $U(1)$-valued field because $\Z_2\Omega$
would appear to leave $A_6$ invariant. Fortunately, this is not the
case. The  $\Z_2\Omega$ action lifts to an M theory orientifold which
acts on the M theory 3-form, $C_3$, by inversion. Tracing the action on
the dual potential shows that this particular component of $A_6$ is
also inverted.

We now dualize along all six circles of the torus giving IIB on, 
$$ T^6/ \Omega (-1)^{F_L} \Z_2, $$
with $A_6$ flux. We can now S-dualize the background. Typically,
commuting S-duality with an orientifold action is problematic. By the
criterion described in~\cite{Sen:1997yy}, S-duality should be permitted in this
situation. This action leaves $ \Omega (-1)^{F_L} \Z_2$ invariant, but
converts $A_6$ to the RR 6-form, $C_6$. We
shall see further evidence for this equivalence when we discuss F
theory compactifications. We can now T-dualize on four circles of the
torus giving IIB on, 
$$ \left( T^2/\Z_2 \times T^4 \right) / \Omega (-1)^{F_L}, $$
with $C_2$ flux through $T^2/\Z_2$. This is a situation which is dual
to a compactification of type I with no vector structure by the
following duality chain: type I on $T^2$ with no vector structure can
be viewed as IIB on $T^2/\Omega$ with $B_2$ flux. Dualizing on both
circles of $T^2$ gives IIB on,
$$ T^2/\Omega (-1)^{F_L} \Z_2,$$
with $B_2$ flux. S-dualizing this configuration gives the desired
orientifold with $C_2$ flux. This chain of reasoning, while not water
tight, suggests that our new orientifold sits in the same moduli
space as a compactification with no vector structure. The rank of the
gauge group is therefore reduced by $8$. 

Two comments are in order: what we currently lack is a way of
characterizing the effect of the
$A_6$ flux. In the case of $B_2$ flux, we could describe the effect in
terms of an exotic Stiefel-Whitney class~\cite{Berkooz:1996iz}, ${\widetilde w_2}$, for the
$\Spin(32)/\Z_2$ bundle. This naturally only involves two-cycles of the
compactification space $\M_6$.

For the case of $A_6$ flux, we need an
analogous statement. A few facts might be helpful toward this end:
unfortunately, the cohomology of $B(\Spin(32)/\Z_2)$ contains no exotic `$w_6$'
class~\cite{Morgan}. Interestingly, however, the cohomology of $B\SO(32)$ does
contain a $w_6$. Since the perturbative type I gauge group is actually
$\O(32)/\Z_2$, the correct characterization of the resulting type I
compactification might well involve an obstruction preventing a lift
of an $\SO$-bundle
to a $\Spin$-bundle. Clearly, a better understanding of the physics of $A_6$ is needed.

As a second comment, we note that in the field theory example
mentioned in the introduction, $\langle \phi\rangle $ could be detected by
monopole 
instantons. These instantons modify BPS couplings in the effective
action, like the $4$ photon
$F^4$ interaction, in a way that can be determined by
supersymmetry~\cite{Paban:1998mp, Dorey:2001ym}. This is not the case in our type I construction. The
natural probes for $\langle \phi\rangle $ are Euclidean NS5-branes wrapping
$\M_6$. However, these configurations are projected out of the spectrum
by the orientifold action. This is the primary reason that the effect
of $A_6$ flux is difficult to determine.

\subsection{Variants of F theory}

If we start with type I on an elliptically-fibered space, we can
imagine T-dualizing along each cycle of the torus fiber to obtain a
IIB orientifold on,
$$ \M_6 /\Omega (-1)^{F_{L}} \Z_2, $$
where the $\Z_2$ acts on both cycles of the fiber. This is the
orientifold limit of F theory compactified on an eight-dimensional
space $\M_8$~\cite{Sen:1997gv}. 

If we start with $A_6$ flux then we end up with $A_6$ flux through
$\M_6$. However, there are two natural discrete fluxes in this
orientifold construction: $A_6$ flux and RR $C_6$ flux through
$\M_6$. Both give $U(1)$-valued scalars which project to $\Z_2$-valued
scalars under the action $\Omega (-1)^{F_{L}} \Z_2$. If we dualize
back to type I along the torus fiber, we see that $C_6$ corresponds to
the possibility of turning on a discrete $C_4$ flux through the base
of the elliptically-fibered $\M_6$. It is natural that both these
possibilities should be on the same footing in an F theory
construction. 

F theory is a just a particular limit of M theory compactified on
$\M_8$ where the volume of the $T^2$ fiber is taken to
zero~\cite{Vafa:1996xn}. How are these discrete choices realized in M
theory? It is easy to trace back the origin of $A_6$ and $C_6$ in type
IIB to M theory. These two potentials arise in the following way: the
metric, $G$, reduced on the $T^2$ fiber gives rise to $2$ Kaluza-Klein
gauge-fields, $A_1$ and $A_2$. 

At the orientifold point, it makes sense to discuss these two
gauge-fields. In three dimensions, we can dualize each of them to a
compact scalar which is $\Z_2$-valued under the orientifold action. 
This construction is reminiscent of the eight-dimensional
F theory dual for the CHL string advocated in~\cite{Bershadsky:1998vn,
Berglund:1998va}. We certainly do not expect all choices for these
$\Z_2$-valued scalars to give physically distinct theories. S-duality will identify
various combinations of scalar field expectation values.  

\subsection{More general constructions}

Our preceeding discussion leads us to a fairly general picture. Let us
consider type II compactified to $d$ space-time dimensions on, 
$$ \M_{10-d}/G.$$ 
For simplicity, let us assume that the covering space, $ \M_{10-d}$,
is smooth. We are interested in the spectrum of $(d-2)$-forms that
arise by compactifying the type II string on the cover. These forms
dualize to scalars in $d$ dimensions with a moduli space, $\M^\phi$.
In general, the group, $G$, will not act freely on $\M^\phi$. Each
disconnected component of $\M^\phi/G$ corresponds to a (classically) distinct
compactification.   

\subsubsection{A type II example}

Let us turn to an example. Consider the type II string
compactified to three dimensions on, 
$$ T^4/\Z_2 \times T^3. $$
Prior to orbifolding by the $\Z_2$, we have a variety of Kaluza-Klein
gauge-fields obtained by reducing the metric $G$ and various
$B$-fields along a one-cycle of $T^4$. Each can be dualized to a
compact scalar in three dimensions. Under the orbifold action, there
are $2$ possible fixed points. 

In this model, there are additional discrete choices
beyond the $4$ $\Z_2$-valued scalars coming from $G$ and the $4$ from
$B_2$. 
For the type IIA
string, there are $12$ additional $\Z_2$-valued scalars obtained by
reducing $C_3$ along a $1$-cycle of $T^4/\Z_2$ and a $1$-cycle of
$T^3$. The RR potential $C_5$ gives $16$ scalars, while $C_7$ gives
$4$ more scalars. The NS potential $A_6$ gives $12$ scalars. 

There is
a final possibility which comes about in the following way: the metric
reduced on a $1$-cycle of $T^3$ gives a $9$-dimensional KK
gauge-field. Dualizing this gauge-field gives a $6$-form potential
$G_6$ which couples to Kaluza-Klein monopoles with respect to the
chosen $1$-cycle. From each of these $3$ $G_6$ potentials, we obtain
$4$ $\Z_2$-valued scalars, giving an additional $12$
possibilities. There may well be more possibilities, but these $64$ are the
simplest $\Z_2$-valued scalars to describe. Again, we note that many of these
choices should give the same low-energy physics.

How does a non-trivial fixed point for one of these scalars change the
low-energy physics? What we expect from arguments given
in~\cite{triples} is a compactification with some frozen
singularities. This is manifest for one case: the $\Z_2$ scalars that
arise from dualizing $C_7$ correspond to a choice of RR 1-form flux
through $T^4/\Z_2$. This is in the class of compactifications
considered in~\cite{Schwarz:1995bj}, and studied in detail
in~\cite{triples}. Of the $16$ $A_1$ singularities on $T^4/\Z_2$, $8$
are frozen by this flux. For more general fluxes, we
again expect frozen singularities though the nature of those
singularities is mysterious. The $1$-form flux is particularly
nice because the resulting compactification is geometric in M
theory. The general case is, unfortunately, not so easy to
analyze. 

\subsubsection{Type I in three dimensions}

Let us now turn to type I on $T^7$. It is natural to expect a set
of discrete choices dual to those we just found in the type II
compactification. We start with IIB on, 
$$ T^7/\Omega. $$
{}From $B_2$ reduced on a $1$-cycle of $T^7$ we obtain $7$
$\Z_2$-valued scalars. $C_4$ reduced on a $3$-cycle gives $35$
choices, while $C_8$ gives one scalar. Lastly, $A_6$ reduced on a
$5$-cycle gives $21$ scalars for a total of $64$
$\Z_2$-valued scalars. 

This is a particularly nice example for the following reason. The
$\Z_2$-valued scalar that arises by reducing $A_6$ on a $5$-cycle and
dualizing can be described a second way. It arises by reducing $B_2$
on the $2$-cycle dual to the $5$-cycle of $T^7$. However,
this $\Z_2$ choice just corresponds to a type I compactification with
(or without) vector structure! We hope that this is compelling
evidence that an $A_6$ background can give a distinct anomaly free
compactification which can be analyzed. 

\subsubsection{M theory on $G_2$ spaces}

The last class of example that we shall examine involves
compactifications of M theory on spaces of $G_2$ holonomy. One of the
simpler cases is not really a $G_2$ compactification, but a `barely
$G_2$' compactification~\cite{Joyce, Harvey:1999as} still preserving
$4$ supersymmetries, 
$$ \left( S^1\times \M_6 \right)/ \Z_2. $$
The $\Z_2$ action acts without fixed points on $\M_6$ but inverts the
circle coordinate. Reducing the M theory $3$-form on $S^1$ gives a
$2$-form which dualizes to a compact scalar. After the $\Z_2$ quotient,
we are left with $2$ fixed points.

This case is more mysterious than
the type II examples because there is no apparent change in the number
of moduli. The difference (if any) between the two possible compactifications
is more subtle than in prior examples. 

A particularly simple way to obtain a (singular) $G_2$
compactification is by starting with a type IIA orientifold. The
strong coupling description will involve M theory on a $G_2$
space.  Some examples of this kind have been recently studied
in~\cite{Blumenhagen:2000ev, Cvetic:2001nr, Kachru:2001je}. Our starting point is IIA on, 
$$ \M_6 / \Omega (-1)^{F_L} G, $$
where $G$ inverts $3$ circles of an appropriately chosen $\M_6$.  
The potential $A_6$ is inverted by this action. In a by now familiar
fashion, we see that there are $2$ classes of IIA orientifold on
$\M_6$. In the corresponding M theory model, we have an $A_6$
background through $\M_6$. This is essentially the `barely $G_2$' case
where we allow an action on $\M_6$ with possible fixed points. The
case of $\M_6 = T^6$ falls into our earlier duality chain, and with
non-trivial $A_6$ flux, we again 
expect a rank reduction of $8$ in the space-time gauge group.

\section{Acknowledgements}
It is our pleasure to thank A. Keurentjes, 
A. Lawrence, J. Morgan, A. Sen, and E. Witten
for helpful   
conversations. The work of D.~R.~M. is supported in part by National
Science Foundation grant DMS-0074072; the work of S.~S. is supported in
part by an NSF CAREER Grant 
No. PHY-0094328, and by the Alfred P. Sloan Foundation. We would
like to thank the Aspen Center for Physics for hospitality during the
completion of this work.

\newpage
%\bibliographystyle{amsunsrt-es}
%\bibliography{myrefs}
%\end{document}
%

\end{document}